# Explaining conflict violence in terms of conflict actor dynamics

Katerina Tkacova[1,2], Annette Idler[1,2] [*], Neil Johnson[3] and Eduardo López[4]

[1] Blavatnik School of Government, University of Oxford, Radcliffe Observatory Quarter, Woodstock Road, Oxford, OX2 6GG, United Kingdom

[2] Global Security Programme, Pembroke College, St. Aldates, Oxford, OX1 1DW, United Kingdom

[3] Dynamic Online Networks Laboratory, George Washington University, Washington D.C., 20052, United States

[4] Department of Computational and Data Sciences, George Mason University, 4400 University Dr, Fairfax, 22030, Virginia, United States

[*] Corresponding author: annette.idler@bsg.ox.ac.uk

*We study the severity of conflict-related violence in Colombia at an unprecedented granular scale in space and across time. Splitting the data into different geographical regions and different historically-relevant periods, we uncover variations in the patterns of conflict severity which we then explain in terms of local conflict actors' different collective behaviors and/or conditions using a simple mathematical model of conflict actors' grouping dynamics (coalescence and fragmentation). Specifically, variations in the approximate scaling values of the distributions of event lethalities can be explained by the changing strength ratio of the local conflict actors for distinct conflict eras and organizational regions. In this way, our findings open the door to a new granular spectroscopy of human conflicts in terms of local conflict actor strength ratios for any armed conflict.*

## Introduction

Since Richardson demonstrated that the distribution of the severity (size) of wars follows an approximate power-law[1,2], other researchers have identified a similar result in phenomena such as protests[3], modern wars[4], insurgencies[5,6], and terrorism[7,8]. Those studies mostly focus on the existence of this distribution across such violent scenarios. However, our knowledge and understanding of the variation of the severity distribution in time and space *within* a given conflict is very limited -- and so too is what that knowledge might then tell us about human behavior within a conflict.

Our aim in this paper is to contribute to this literature by exploring the severity of conflict events within the armed conflict in Colombia at an unprecedented granular scale in space and across time. Moreover, we split the conflict into specific segments that all have historical and/or organizational meaning, i.e. smaller geographical units and three periods that represent distinct eras of the conflict[9]. We then test the fitting of event severity data within these separate space-time sections of the armed conflict, to a power-law (i.e. scale-free) distribution. Combining this with a previously published mathematical model of conflict actor group behavior[5], this allows us to interpret the evolution of the conflict across time and space. Specifically, because of this unique granular focus, we are able to interpret the patterns in this important but complex conflict in terms of a spatial-temporal decomposition guided by an established multi-actor conflict framework[9] where the conflict is best understood as a set of interrelated but approximately separable subconflicts.

As background, we note that a power-law distribution describes the frequency of conflict events of a size $x$, as $p(x) = x^{-\alpha}/\zeta(\alpha, x_{\min})$, where $\alpha$ is a scaling exponent and $\zeta(\alpha, x_{\min})$ is a Hurwitz zeta function for discrete power-laws which is more appropriate for our type of data[8,10,11]. When plotted on a log-log plot where the x-axis is the logarithm of the event size and the y-axis is the is the logarithm of the cumulative probability $P(x)$ of that event being at least size $x$, the data produce a straight line with a slope $\alpha - 1$ [12] (N.B. all figures show P(x)). A power-law distribution is also often called scale-free. This means that increasing the scale or changing the unit in which $x$ is measured does not result in any change in the shape of the distribution [8].

Various mathematical mechanisms produce power-law distributions in physics, biology, and social systems [14]. For instance, forest fires[15], wars[4] and protests[3], and the collapse of civilizations[16] are modeled according to self-organized criticality (SOC) suggesting that large composite systems tend to reach a critical state in which even small events can cause a catastrophe, such as a landslide or a volcano eruption[17]. Although widely used, SOC does not consider the interaction between two or more actors. Thus, to explain the existence of power-law distributions, we rely on a published coalescence-fragmentation model for conflict actor grouping dynamics[18] that allows us to include armed groups as dynamically evolving clusters of populations that can lose or gain their members over time, i.e., fragment or coalesce. The coalescence-fragmentation model was further developed to account for two or more opposing sides of conflict[5]. This model suggests a possible range of $\alpha$, the scaling exponent for the power-law distribution, conditioned by the varying strength ratio of conflict actors. It means that interactions of conflict actors yield distinct values of $\alpha$ in the individual regions that give insights into the power ratio between relevant actors.

To give an overview of how this coalescence-fragmentation model behaves, we focus on a hypothetical situation with two conflict actors: government and insurgents. The conflict actors have a total strength that consists of the combined populations of $N_a$ (government) and $N_b$ (insurgents). Both populations consist of fighters (agents) who form armed groups (clusters) via the coalescence and fragmentation processes that operate as follows. First, two agents are selected from the combined populations $N_a$ and $N_b$. If they are from the same side of the conflict, for instance, both agents belong to the state forces, the clusters of which they are members coalesce, i.e. the two armed groups join forces. If the selected agents are from opposing sides, they fight, resulting in casualties on both sides. Scales for smaller and larger groups engaged in the violent encounter, denoted as $C_S$ and $C_L$, respectively, determine the number of losses. These scales function similarly to the attrition coefficient in Lanchester's equations [19,20]. Hence, the smaller armed group fragments into even smaller groups of fighters after the battle. The frequency of events and their size, defined as combined losses of groups engagement in battle, yield approximate power-law distribution with scaling exponent $\alpha = 2.5$. The obtained value of $\alpha$ is in line with other research fitting data from various conflicts involving insurgencies in the power-law distribution and solving the coalescence-fragmentation model analytically[5].

The ratio between the strength of the conflict sides affects the value of the scaling exponent $\alpha$[5]. Weaker sides tend to create smaller groups on average than the stronger side due to the larger number of opposing agents within the population. Assuming the weaker side suffers higher losses than the stronger side, the greater difference between the strength of the two enemy populations leads to the steeper slope of the exponent $\alpha$, meaning a decreasing probability of large battles. Thus the exponent $\alpha$ can serve as a proxy for the strength ratio between conflict sides. By studying the values of $\alpha$, we can better understand the strength ratio between conflict actors and thereby provide important insights into a given conflict's dynamics.

Here we examine the within-conflict variation in $\alpha$ for the case of the Colombian armed conflict in which violence has fluctuated over time and across space. The conflict began with an episode of violence, "La Violencia", in the late 1940s and turned into a civil war between leftist guerrillas, including the then largest non-state armed actor, the Revolutionary Armed Forces of Colombia (FARC), founded in 1964, and the government. Soon paramilitary groups formed to fight the guerrillas and protect large landowners. With the rise of cocaine production in the 1970s and 1980s in the country, drug cartels and other criminal actors became involved in the conflict. After the FARC's demobilization in 2017, multiple violent non-state actors continued to engage in violence[9,21].

The Colombian armed conflict manifests remarkable changes over its long and violent history; therefore, it represents an ideal case for our study. The intensity and frequency of violence has been fluctuating over time and across space, especially, during the demobilization of the paramilitaries between 2003 and 2006, following the peace agreement with the largest non-state armed actor FARC in 2016, and in the run-up to most of the recent presidential elections, which often introduced changes in the course of governmental strategies to tackle insurgent groups.

Measuring these variations within a given conflict is important because it highlights the interdependent and interconnected character of conflict actors across different subconflicts that together form a larger conflict. It is possible, for example, that in a large conflict the measurements of local events in that conflict share the same features as the conflict as a whole (an expectation that would come from the scale-free nature of conflict statistics). This has implications for strategic and policy decisions on how to understand the overall conflict, and how to prioritize solutions to it. Our work offers a first glimpse into these different scenarios.

# Results

## Identifying the conflict segments

We conceptualize the armed conflict in Colombia as a multi-actor conflict with the involvement of the following actors in the period 1989 – 2018: state forces, left-wing guerrillas, paramilitaries, and criminal groups involved in conflict-related violence[9]. We split the conflict into three periods, each featuring a distinct government strategy toward the conflict actors reflecting the change in the strength of the main actors. In addition, we identify the distinct regions within the Colombian armed conflict based on organizational structure and mutual interactions of the main conflict actors. Through this approach we obtain regions that allow us to estimate the strength ratio of the regional branches or sections of the main conflict actors.

The first period (1989-1999) comprises various presidencies characterized by peace negotiations with various guerrilla groups. The peace process initiated by former President Samper failed and left FARC rebels strengthened[22,23]. The second period (2000-2009) represents former President Uribe's administration. His hard-line Democratic Security Policy sought to weaken the FARC through the counternarcotics and counterinsurgency strategy "Plan Colombia" that was supported by the United States. The third one (2010-2018) covers 2010 when President Santos took office. He later initiated peace talks with the FARC, followed by signing the peace deal between the government and these rebels in 2016 and the FARC's subsequent demobilization in 2017[24].

Historically, and to a great extent influenced by geography, Colombia's economic, social, and political dynamics have significantly varied across regions, reflected in distinct cultures and identities and influential regional elites[25]. Similarly, the armed conflict patterns exhibit regional diversity[26]. Many conflict actors' organizational structures are adjusted to the primary geographical location in which they are active; these locations coincide across key conflict actors, including the Colombian National Police, the National Army, and the FARC. For instance, the 1st division of the National Army covers approximately the same area as region 8 of the National Police and FARC's Caribbean Bloc. Given the geographical proximity, we assume that the 1st division and police units from region 8 are more likely to interact with the Caribbean Bloc than other FARC blocs and vice versa. Hence, we assume that there are more interactions between these three than Army divisions, police regions, and blocs active in other parts of Colombia and that the interactions between the 1st division, police units from region 8, and the Caribbean Bloc are to some extent independent from the other divisions, police regions, and FARC blocs.

Given that over 75% of the interactions in our data are between the FARC and the state forces (see Table 1), we construct the spatial split of the Colombian armed conflict based on the geographical overlap of the institutional structure of the National Police, the National Army, and the FARC. Although we do not use the structures of other conflict actors to construct the conflict segments, we include them when analyzing the conflict events. In this way, we can isolate the local strength ratio of the actors.

| Conflict actor | Conflict events involvement (N) |
|---|---|
| Government of Colombia (the National Police and the National Army combined) | 3164 |
| FARC | 2595 |
| ELN | 625 |
| AUC | 130 |
| EPL - Megateo | 36 |
| FARC dissidents | 28 |
| EPL | 10 |
| Medellin Cartel | 8 |
| Cali Cartel | 4 |
| PEPES | 4 |
| Bloque Central Bolivar | 1 |
| ELN, FARC | 1 |

Table 1: Number of conflict events per actor for the period 1989-2018. Data obtained from the UCDP GED.

We proceed by first determining the overlap between the eight National Police regions and the National Army divisions and form seven resulting regions. Both organizations mirror each other's spatial organization with two exceptions. The 3$^{rd}$ division of the National Army covers an area of regions 3 and 4 of the National Police. Similarly, combining the 4$^{th}$ and 8$^{th}$ National Army divisions, we cover the area of region 7 of the National Police. Second, we identify the primary locations where the FARC blocs are active and overlay them with the regions identified in the previous step. Although the FARC blocs are to some extent mobile and do not fully spatially coincide with the state forces, their geographical overlap still represents a reasonable approximation to identify the resulting regions. To match the National Police and National Army organizational structure across Colombia, we split FARC's North Western Bloc into regions 2 and 3.

Similarly, while region 5 comprises mainly FARC's Central Bloc, we also added parts of the Eastern Bloc and the Southern Bloc. In both cases, our decision to split some FARC Blocs was also driven by the natural geographical features, including elevation and terrain type forming, in the case of Colombia, prominent boundaries of the armed conflict on its own (see Figure 1). To compile the regions based on the organizational structure of the main conflict actors, we consulted several sources[27–33].

Colombia's Constitution of 1991 (article 287)[34] decentralized the country, granting departments some autonomy to govern themselves and administer and use the resources they are allocated[35]. Conflict dynamics vary across departments and regions—the departments' administrative autonomy and the somewhat decentralized nature of the police forces contribute to this diversity. During former President Uribe's administration, for example, the national government considered the FARC a terrorist group and acted accordingly, including refusing to engage in talks with the FARC, whereas the (regional) government of the Nariño department engaged in peacebuilding activities open to dialogue[36]. We, therefore, include the analysis of the individual departments in addition to the regions described above.

| Region | National Army | National Police | FARC Blocs | Departments/administrative units |
|---|---|---|---|---|
| Region 1 | 1st Division | Region 8 | Caribbean Bloc | La Guajira, Cesar, Magdalena, Atlántico, Bolivar, Sucre |
| Region 2 | 7th Division | Region 6 | North Western Bloc | Córdoba, Antioquia, Chocó |
| Region 3 | 2nd Division | Region 5 | North Western Bloc | Norte Santander, Santander |
| Region 4 | 3rd Division | Region 3 and 4 | Western Bloc | Nariño, Cauca, Valle de Cauca, Risaralda, Caldas, Quindio |
| Region 5 | 5th Division | Region 1 | Part of the Eastern Bloc, Central Bloc and part of the Southern Bloc | Cundinamarca, Boyacá, Tolima, Huila, Bogota |
| Region 6 | 6th Division | Region 2 | Southern Bloc | Caquetá, Putumayo, Amazonas |
| Region 7 | 4th and 8th Divisions | Region 7 | Eastern Bloc | Arauca, Casanare, Meta, Vichada, Guainia, Guaviare, Vaupes |

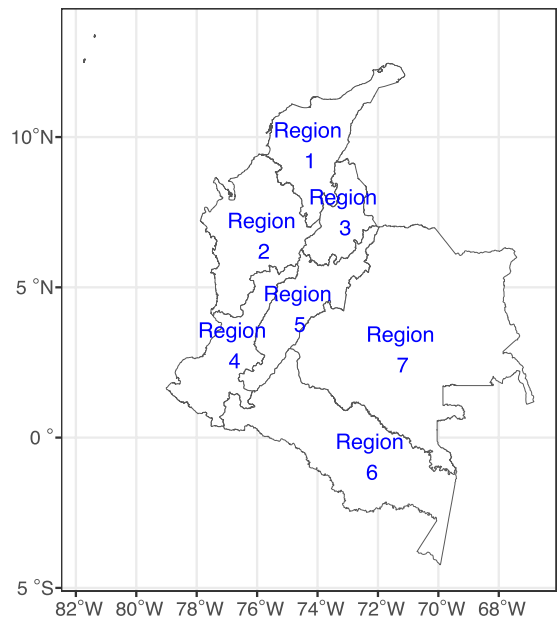

Figure 1: Regions compiled by overlaying the organizational structure of the National Police, the National Army and FARC Blocs. Top: Table listing the National Army divisions, the National Police regions, the FARC blocs and departments for each compiled region. Bottom: Map showing the geographical distribution of the compiled regions.

Data

We obtain data from the Uppsala Conflict Data Programme Georeferenced Event Dataset version 19.1 (UCDP GED) [37,38] that provides information on geo-referenced conflict events and is widely used by conflict researchers [39–41].

Following the conflict conceptualization as a "dynamic multi-actor setting of organized violence with one or more contested issues between two or more conflict actors resulting in deaths"[42], we identify the relevant conflict actors and conflict events for our analysis. Accordingly, the actors selected from the UCDP GED data are the Colombian government (forces representing the Colombian government, including the National Police and the National Army), the FARC, FARC dissidents, the National Liberation Army (ELN), the Popular Liberation Army (EPL), EPL – Megateo, the United Self-Defences of Colombia (AUC), the Central Bolivar Bloc, the "Persecuted by Pablo Escobar" (PEPES), the Cali cartel, and the Medellin cartel. We include conflict events involving the selected conflict actors that resulted in at least one battle-related death. Conflict events that cannot be assigned to specific departments due to their low geo-precision are excluded from our analysis. Based on these selection criteria, we obtained 3,303 events for 1989-2018. We split data according to the event's location and date to obtain data for the individual segments of the conflict.

Findings

This section describes the results obtained after attempting to fit our data for the individual segments of the armed conflict in Colombia to the power-law distribution. We proceed as follows to test whether our data have power-law distributions. First, we estimate $\alpha$ and $x_{min}$ from our data to draw ideal power-law distribution based on these parameters. Second, we calculate the Kolmogorov-Smirnov goodness of fit test and obtain the P-value via the bootstrap procedure with 5000 iterations. We accept that data follow an approximate power-law distribution if the resulting P-value is equal to or greater than 0.1[10].

The value of $x_{min}$ sets the threshold for the minimum size events, in our case number of fatalities, for the power-law to apply. In other words, conflict events with a number of fatalities smaller than $x_{min}$ are not part of the data forming the power-law distribution. The estimated value of the exponent $\alpha$ suggests a possible level of imbalance between the strength of the conflict actors. Based on the coalescence-fragmentation model[5], a higher value of $\alpha$ means that larger battles are less frequent; therefore, the strength of the conflict actors is more asymmetric. While the scaling exponent describes violence patterns across sections of conflicts (regions and departments in different periods), the P-value confirms the presence of the power-law distribution.

Carrying out this power-law fitting procedure for all of our data over all the studied periods of the Colombian conflict (1989-2018), we obtain an $\alpha$ exponent with a value of 2.54 and P value of the Kolmogorov-Smirnov goodness of fit test greater than 0.1. This confirms existing research that found a power-law distribution with $\alpha$ near 2.5 for modern insurgencies[5] and it confirms that the conflict as a whole follows such a power-law distribution. Figure 2 shows how the size (severity) distribution of conflict events indeed forms an approximate straight line when displayed on a log-log plot where the horizontal axis is the event size, and the vertical axis is the cumulative probability of that event being at least that size. The mean size of conflict events during the studied period is 6.34.

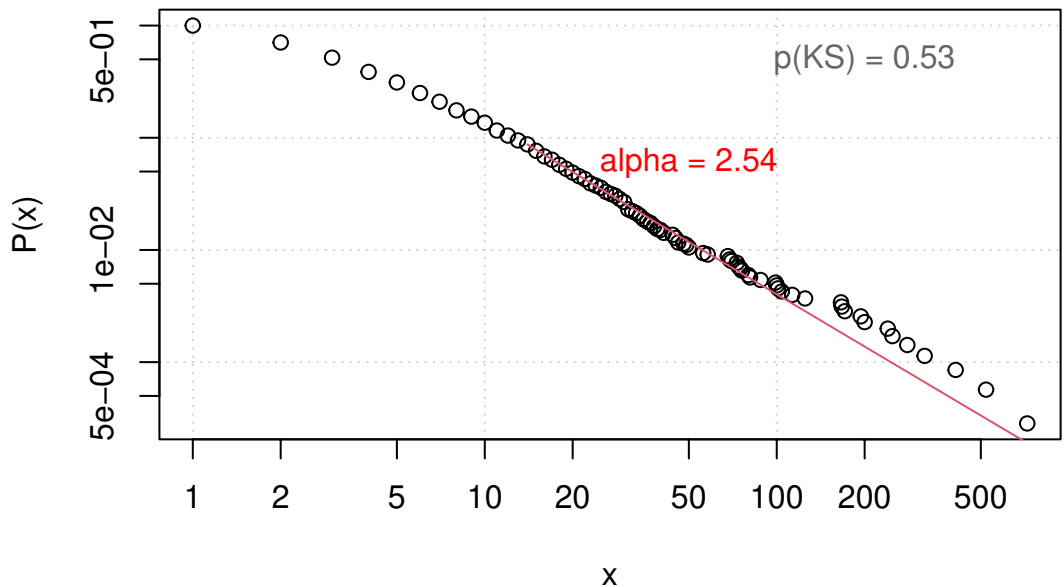

Figure 2: Frequency-size distribution of conflict events in the armed conflict in Colombia, 1989-2018.

*Segments of armed conflict: Regions*

The results for the regions formed by overlaying the institutional structure of the National Police, the National Army, and FARC show that splitting the data into smaller segments based on the predominant mutual interaction of the main actors also follow approximate power-law distributions. The Kolmogorov-Smirnov goodness of fit test yields P-values greater or equal to 0.1 for most of the regions (see complete results in Table 2). Since the data is now being split, there are less data points within a given region than overall, hence the data can appear more noisy, i.e. there is generally more scatter around the straight line.

| Region | Period | Total N of events | N of events in tail | X(min) | Alpha | P-value (KS) |
|---|---|---|---|---|---|---|
| Region 1 | All years | 429 | 258 | 2 | 2.07 | 0.07 |
| Region 1 | 1989-1999 | 91 | 38 | 5 | 2.43 | 0.06 |
| Region 1 | 2000-2009 | 328 | 174 | 2 | 2.17 | 0.77 |
| Region 1 | 2010-2018 | 10 | 8 | 2 | 2.45 | 0.81 |
| Region 2 | All years | 677 | 350 | 3 | 2.00 | 0.00 |
| Region 2 | 1989-1999 | 177 | 34 | 13 | 2.80 | 0.54 |
| Region 2 | 2000-2009 | 447 | 199 | 3 | 2.04 | 0.28 |
| Region 2 | 2010-2018 | 53 | 53 | 1 | 1.79 | 0.04 |
| Region 3 | All years | 305 | 154 | 3 | 2.37 | 0.08 |
| Region 3 | 1989-1999 | 98 | 69 | 3 | 2.36 | 0.18 |
| Region 3 | 2000-2009 | 179 | 122 | 2 | 2.27 | 0.23 |
| Region 3 | 2010-2018 | 28 | 9 | 5 | 3.25 | 0.13 |
| Region 4 | All years | 564 | 97 | 7 | 2.67 | 0.26 |
| Region 4 | 1989-1999 | 78 | 23 | 7 | 3.03 | 0.67 |
| Region 4 | 2000-2009 | 401 | 99 | 5 | 2.46 | 0.86 |
| Region 4 | 2010-2018 | 85 | 42 | 3 | 2.35 | 0.27 |
| Region 5 | All years | 537 | 115 | 6 | 2.75 | 0.21 |
| Region 5 | 1989-1999 | 144 | 47 | 6 | 3.28 | 0.24 |
| Region 5 | 2000-2009 | 365 | 163 | 3 | 2.29 | 0.49 |
| Region 5 | 2010-2018 | 28 | 19 | 2 | 2.35 | 0.01 |
| Region 6 | All years | 302 | 120 | 4 | 2.29 | 0.14 |
| Region 6 | 1989-1999 | 49 | 16 | 9 | 2.51 | 0.59 |
| Region 6 | 2000-2009 | 224 | 56 | 5 | 2.61 | 0.53 |
| Region 6 | 2010-2018 | 29 | 9 | 9 | 5.34 | 0.79 |
| Region 7 | All years | 483 | 38 | 20 | 3.18 | 0.81 |
| Region 7 | 1989-1999 | 139 | 98 | 3 | 1.92 | 0.46 |
| Region 7 | 2000-2009 | 271 | 97 | 5 | 2.39 | 0.25 |
| Region 7 | 2010-2018 | 73 | 20 | 9 | 3.16 | 0.93 |

Table 2: Results obtained from the bootstrapping procedure for regions.

Figure 3 shows that conflict event data for the individual regions mostly follow an approximate power-law distribution except for region 1 in 1989-1999 and regions 2 and 5 in 2010-2018. Notably, all regions in 2000-2009, the most intensive period of the conflict with regard to the number of conflict events, reached the P-value above 0.1. This suggests that the small number of observations might be driving some P-values below the 0.1 thresholds. For example, region 5 in 2010-2018 contains only 28 data points.

The variation of the exponent $\alpha$ provides insights into the patterns of conflict-related violence across time and space. In 1989-1999, $\alpha$ tends to oscillate around 2.5, suggesting a similar pattern of the events' occurrence conditioned by their size in most parts of Colombia, except from regions 4, 5 and 7. During this period, neighbouring regions 4 and 5 in central and west Colombia, with exponents $\alpha = 3.0$ and $\alpha = 3.3$, respectively, experienced large events less often than region 7 in South-East Colombia, with exponent $\alpha = 1.9$. In the second period, 2000-2009, the $\alpha$ values decrease and drop below 2.5 except for region 6 in southern Colombia. Thus, we observe relatively larger events more often across most parts of Colombia compared to the previous period. Results for the period 2010-2018 point to more significant variation in the conflict dynamics across the regions. For example, region 6, located in central Colombia with exponent $\alpha = 5.3$, experienced predominantly small events. Similarly, regions 7 and 3, located in eastern Colombia at the border with Venezuela, had relatively small events. Compared to other regions, larger events were more frequent in region 4 situated at the borders with Ecuador, as its exponent $\alpha$ is 2.4.

According to the coalescence-fragmentation model[5], the exponent $\alpha$ variation is based on the strength ratio of the conflict actors engaged in fighting. Thus the results described above can provide insights into conflict dynamics that go beyond the frequency of conflict events conditioned by their size. Relatively small differences in the exponent $\alpha$ across most of Colombia in 1989-1999 suggest a similar strength ratio of conflict actors across Colombia. At that time, the guerrillas and other non-state conflict actors profited greatly from the illegal drug trade. The FARC, for example, who controlled land for coca cultivation, required drug traffickers to pay for establishing laboratories to process coca into cocaine. The FARC's large income, hierarchically centralized structure, and ability to govern territory and its inhabitants[43] explain the weak position of the Colombian state relative to FARC and the lack of variation in strength ratio between those two actors across most of Colombian territory. The state offensive after 2000 led to intensified violence, forced the guerrillas to Colombia's geographical margins, and gradually weakened them. Smaller values of the exponent $\alpha$ in most regions hint at the lesser strength disparity and a higher frequency of large events compared to the previous period. During 2000-2009, the Colombian government increased its military expenditure from $2.6 billion in 2001 to $11 billion in 2010. The pressure from the Colombian government led to more strength parity as the FARC weakened. The government grew stronger[43] and matched the power of many non-state conflict actors benefiting from the illegal drug trade. The period 2010-2018 saw a de-escalating trend. The weakened FARC lost territory across Colombia, resulting in a power vacuum filled by other non-state conflict actors. However, none of these non-state actors managed to attain as strong a presence across most regions as the FARC did in previous periods. As our results suggest, the strength ratio between the government and non-state actors varied across Colombia after 2010.

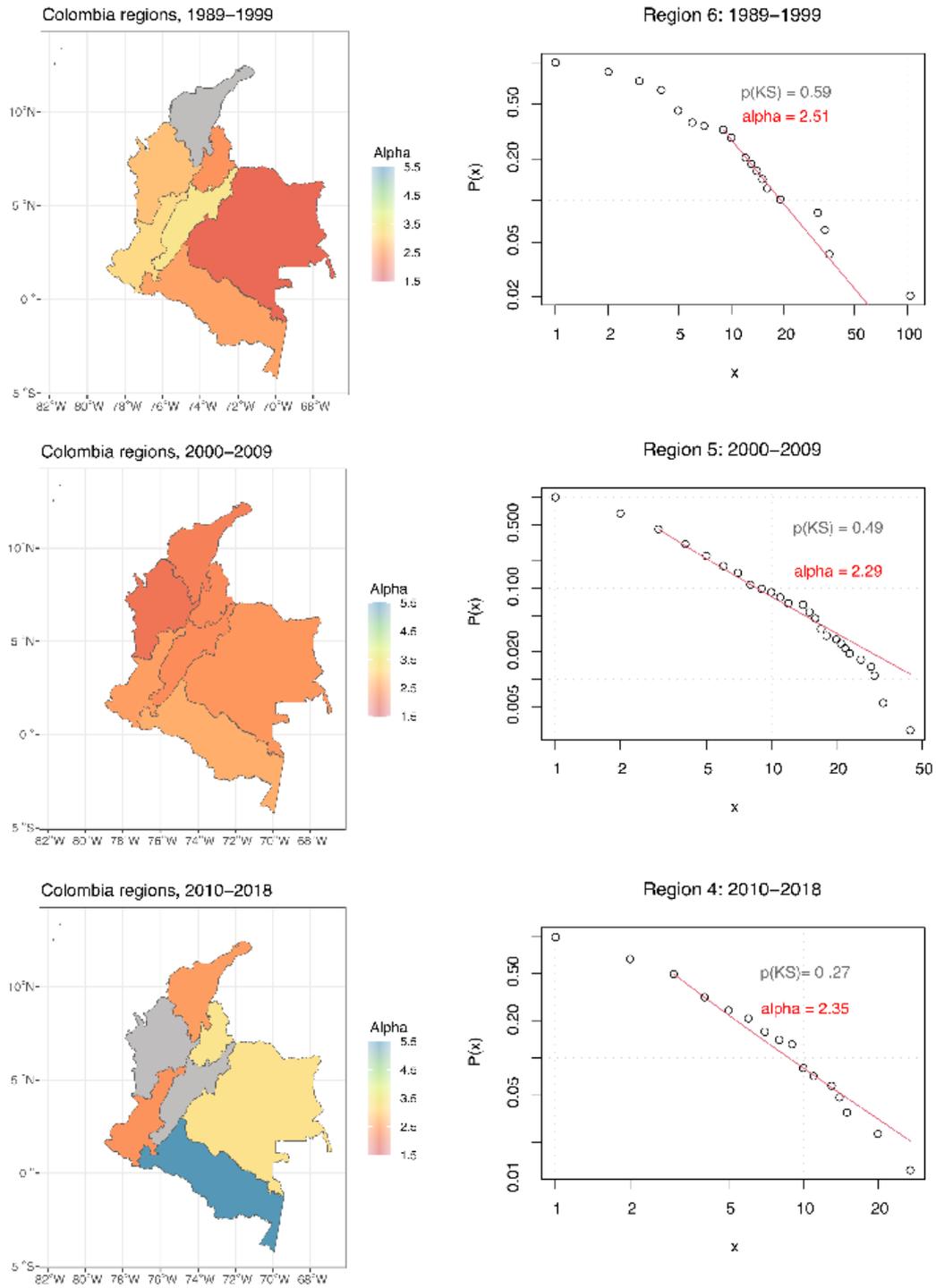

Figure 3: Results for the regions in Colombia obtained by fitting the data in the power-law distribution and calculating the Kolmogorov-Smirnov goodness of fit test. Left: Geographical distribution of exponent alpha, regions with P-value < 0.1 in grey. Right: CCDF for the selected regions.

*Segments of armed conflict: Departments*

Figure 4 shows the results obtained when we divided the data into even smaller sections of armed conflict. As explained above, Colombian departments possess some level of autonomy that can manifest in their approach to non-state actors. Thus it is reasonable to assume that fitting the conflict event data for the individual departments might provide further insights into the conflict dynamics. The P-value obtained by carrying out the Kolmogorov-Smirnov goodness of fit is mostly above the 0.1 threshold (see complete results in Tables 3a and 3b), justifying our decision to analyze also the smaller segments of armed conflict. Focusing on the smaller conflict segments allows us to better interpret our results in the context of the local conflict dynamics.

A large number of diverse conflict actors in Colombia might affect the scaling exponents for individual departments. Assuming that the presence of more actors leads to fragmentation of resources and greater competition for local support and recruits, we expect departments with more actors to have a larger imbalance of strength as the populations from which armed actors can form their fighting units are smaller relative to the governmental forces due to the more intense competition for support from civilians. Therefore, larger events occur less often.

The Arauca department experienced larger battles more often than many other departments in the 1990s, suggesting that the strength of the main conflict actors, namely the Colombian government, the FARC, and the ELN, in this department, was more balanced. This changed in the second period starting in 2000 when paramilitaries joined the fight against the guerrillas resulting in less frequent large battles. The trend continued in the third period beginning in 2010, when the scaling exponent reached the value of 3.81, suggesting that non-state actors became weaker relative to the state forces. Certainly, the demobilization of the paramilitaries and later the FARC played a role as strong conflict actors were replaced by smaller conflict actors such as FARC dissidents. We observe a similar trend in the Casanare department, where larger battles became less frequent over time.

| Department / political unit | Period | Total N of events | N of events in tail | X(min) | Alpha | P-value (KS) |
|---|---|---|---|---|---|---|
| Antioquia department | All years | 564 | 280 | 3 | 2.06 | 0.03 |
| Antioquia department | 1989-1999 | 147 | 28 | 13 | 2.98 | 0.77 |
| Antioquia department | 2000-2009 | 379 | 159 | 3 | 2.16 | 0.36 |
| Antioquia department | 2010-2018 | 38 | 38 | 1 | 1.82 | 0.34 |
| Arauca department | All years | 146 | 36 | 7 | 2.60 | 0.44 |
| Arauca department | 1989-1999 | 35 | 22 | 4 | 2.26 | 0.21 |
| Arauca department | 2000-2009 | 78 | 21 | 5 | 3.15 | 0.91 |
| Arauca department | 2010-2018 | 33 | 11 | 10 | 3.81 | 0.97 |
| Bogotá | All years | 53 | 14 | 5 | 2.92 | 0.77 |
| Bogotá | 1989-1999 | 24 | 16 | 2 | 2.20 | 0.12 |
| Bogotá | 2000-2009 | 25 | 25 | 1 | 1.72 | 0.05 |
| Bolívar department | All years | 129 | 83 | 2 | 1.86 | 0.12 |
| Bolívar department | 1989-1999 | 31 | 28 | 2 | 1.76 | 0.01 |
| Bolívar department | 2000-2009 | 92 | 92 | 1 | 1.76 | 0.18 |
| Boyacá department | All years | 74 | 55 | 2 | 1.99 | 0.07 |
| Boyacá department | 1989-1999 | 21 | 8 | 8 | 4.19 | 0.14 |
| Boyacá department | 2000-2009 | 51 | 24 | 3 | 2.55 | 0.12 |
| Córdoba department | All years | 35 | 35 | 1 | 1.52 | 0.39 |
| Córdoba department | 1989-1999 | 11 | 7 | 6 | 1.97 | 0.33 |
| Córdoba department | 2000-2009 | 22 | 22 | 1 | 1.64 | 0.07 |
| Caldas department | All years | 53 | 53 | 1 | 1.94 | 0.02 |
| Caldas department | 2000-2009 | 46 | 46 | 1 | 2.01 | 0.22 |
| Caquetá department | All years | 211 | 63 | 5 | 2.56 | 0.45 |
| Caquetá department | 1989-1999 | 28 | 19 | 3 | 2.03 | 0.18 |
| Caquetá department | 2000-2009 | 160 | 41 | 5 | 2.86 | 0.05 |
| Caquetá department | 2010-2018 | 23 | 8 | 9 | 4.95 | 0.86 |
| Casanare department | All years | 61 | 30 | 3 | 2.34 | 0.77 |
| Casanare department | 1989-1999 | 27 | 15 | 3 | 2.07 | 0.59 |
| Casanare department | 2000-2009 | 29 | 10 | 4 | 3.29 | 0.94 |
| Cauca department | All years | 213 | 38 | 7 | 2.79 | 0.90 |
| Cauca department | 1989-1999 | 29 | 9 | 7 | 3.58 | 0.70 |
| Cauca department | 2000-2009 | 133 | 29 | 6 | 2.56 | 0.64 |
| Cauca department | 2010-2018 | 51 | 22 | 3 | 2.49 | 0.81 |
| Cesar department | All years | 86 | 63 | 2 | 2.33 | 0.19 |
| Cesar department | 1989-1999 | 29 | 24 | 2 | 2.26 | 0.37 |
| Cesar department | 2000-2009 | 55 | 38 | 2 | 2.36 | 0.49 |
| Chocó department | All years | 78 | 51 | 3 | 1.90 | 0.02 |
| Chocó department | 1989-1999 | 19 | 12 | 4 | 2.20 | 0.82 |
| Chocó department | 2000-2009 | 46 | 18 | 8 | 2.23 | 0.49 |
| Chocó department | 2010-2018 | 13 | 5 | 4 | 3.59 | 0.58 |
| Cundinamarca department | All years | 115 | 36 | 5 | 2.62 | 0.66 |
| Cundinamarca department | 1989-1999 | 48 | 19 | 5 | 2.51 | 0.27 |
| Cundinamarca department | 2000-2009 | 66 | 37 | 3 | 2.53 | 0.65 |
| Guaviare department | All years | 50 | 35 | 3 | 2.06 | 0.68 |
| Guaviare department | 1989-1999 | 15 | 11 | 3 | 1.88 | 0.61 |
| Guaviare department | 2000-2009 | 24 | 15 | 4 | 2.27 | 0.56 |
| Guaviare department | 2010-2018 | 11 | 6 | 3 | 2.86 | 0.33 |
| Huila department | All years | 126 | 26 | 6 | 3.38 | 0.48 |
| Huila department | 1989-1999 | 30 | 12 | 6 | 4.99 | 0.47 |
| Huila department | 2000-2009 | 84 | 38 | 3 | 2.49 | 0.85 |
| Huila department | 2010-2018 | 12 | 12 | 1 | 1.92 | 0.11 |
| La Guajira department | All years | 34 | 16 | 3 | 2.29 | 0.47 |
| La Guajira department | 2000-2009 | 30 | 11 | 4 | 2.58 | 0.94 |
| Magdalena department | All years | 105 | 105 | 1 | 2.12 | 0.05 |
| Magdalena department | 1989-1999 | 11 | 6 | 4 | 2.86 | 0.91 |
| Magdalena department | 2000-2009 | 94 | 94 | 1 | 2.25 | 0.04 |

Table 3a (part 1): Results obtained from the bootstrapping procedure for departments.

| Department / political unit | Period | Total N of events | N of events in tail | X(min) | Alpha | P-value (KS) |
|---|---|---|---|---|---|---|
| Meta department | All years | 199 | 22 | 17 | 3.18 | 0.84 |
| Meta department | 1989-1999 | 53 | 19 | 7 | 2.55 | 0.72 |
| Meta department | 2000-2009 | 123 | 22 | 12 | 2.82 | 0.60 |
| Meta department | 2010-2018 | 23 | 12 | 5 | 2.24 | 0.44 |
| Nariño department | All years | 131 | 100 | 2 | 2.09 | 0.05 |
| Nariño department | 1989-1999 | 13 | 10 | 3 | 2.07 | 0.19 |
| Nariño department | 2000-2009 | 93 | 68 | 2 | 2.19 | 0.63 |
| Nariño department | 2010-2018 | 25 | 5 | 9 | 3.76 | 0.57 |
| Norte de Santander department | All years | 180 | 21 | 9 | 3.41 | 0.89 |
| Norte de Santander department | 1989-1999 | 46 | 30 | 3 | 2.17 | 0.14 |
| Norte de Santander department | 2000-2009 | 108 | 74 | 2 | 2.32 | 0.15 |
| Norte de Santander department | 2010-2018 | 26 | 9 | 5 | 3.25 | 0.15 |
| Putumayo department | All years | 89 | 68 | 2 | 1.88 | 0.21 |
| Putumayo department | 1989-1999 | 21 | 17 | 3 | 1.81 | 0.09 |
| Putumayo department | 2000-2009 | 63 | 44 | 2 | 2.03 | 0.68 |
| Quindío department | All years | 11 | 7 | 3 | 2.56 | 0.30 |
| Quindío department | 2000-2009 | 10 | 6 | 3 | 2.65 | 0.34 |
| Risaralda department | All years | 35 | 35 | 1 | 1.94 | 0.72 |
| Risaralda department | 2000-2009 | 31 | 31 | 1 | 2.05 | 0.37 |
| Santander department | All years | 125 | 70 | 3 | 2.51 | 0.43 |
| Santander department | 1989-1999 | 52 | 39 | 3 | 2.55 | 0.75 |
| Santander department | 2000-2009 | 71 | 31 | 3 | 2.46 | 0.72 |
| Sucre department | All years | 69 | 32 | 3 | 2.39 | 0.78 |
| Sucre department | 1989-1999 | 15 | 12 | 3 | 2.02 | 0.51 |
| Sucre department | 2000-2009 | 54 | 33 | 2 | 2.39 | 0.10 |
| Tolima department | All years | 169 | 76 | 3 | 2.22 | 0.14 |
| Tolima department | 1989-1999 | 21 | 14 | 3 | 2.80 | 0.77 |
| Tolima department | 2000-2009 | 139 | 85 | 2 | 2.00 | 0.27 |
| Valle del Cauca department | All years | 121 | 72 | 3 | 2.14 | 0.24 |
| Valle del Cauca department | 1989-1999 | 24 | 22 | 2 | 2.02 | 0.95 |
| Valle del Cauca department | 2000-2009 | 88 | 38 | 4 | 2.23 | 0.73 |
| Vichada department | All years | 14 | 4 | 20 | 5.88 | 0.66 |

Table 3b (part 2): Results obtained from the bootstrapping procedure for departments.

The landlocked Antioquia department experienced the opposite trend. In the 1990s, this department was fragmented with predominantly smaller battles and many actors, including the Colombian government, the FARC, the ELN, the EPL, the paramilitaries, and the Medellin and Cali cartels. The change started in the second period when the number of actors was reduced to the FARC, the ELN, the paramilitaries, and the Colombian government, and large battles became more frequent. This trend continued in the third period. The Cauca department went through similar changes in the frequency of large battles.

This brief description demonstrates how the variation of the patterns of conflict-related violence is associated with the changes in the actor constellations, including the number of actors involved and their relative strength. Such dynamics are often overlooked when only studying the number of conflict events or battles-related deaths. By dividing the armed conflict into three distinct periods and geographically defined segments of conflict, the analysis demonstrates the importance of understanding armed conflicts as fluid, constantly evolving rather than static phenomena. For instance, the change in Casanare from conflict with frequent large battles in the period 1989-1999 to conflict with predominantly small battles in the period 2000-2009 would be missed as both periods experienced a similar number of battles (27 and 29, respectively). Similarly, data on the number of fatalities in Cauca in the second and third periods shows a significant drop from 615 to 174 and, therefore, a decrease in the intensity of violence. Yet, the structure of the conflict events conditioned by their size remained the same, suggesting an unchanged strength ratio of the actors involved.

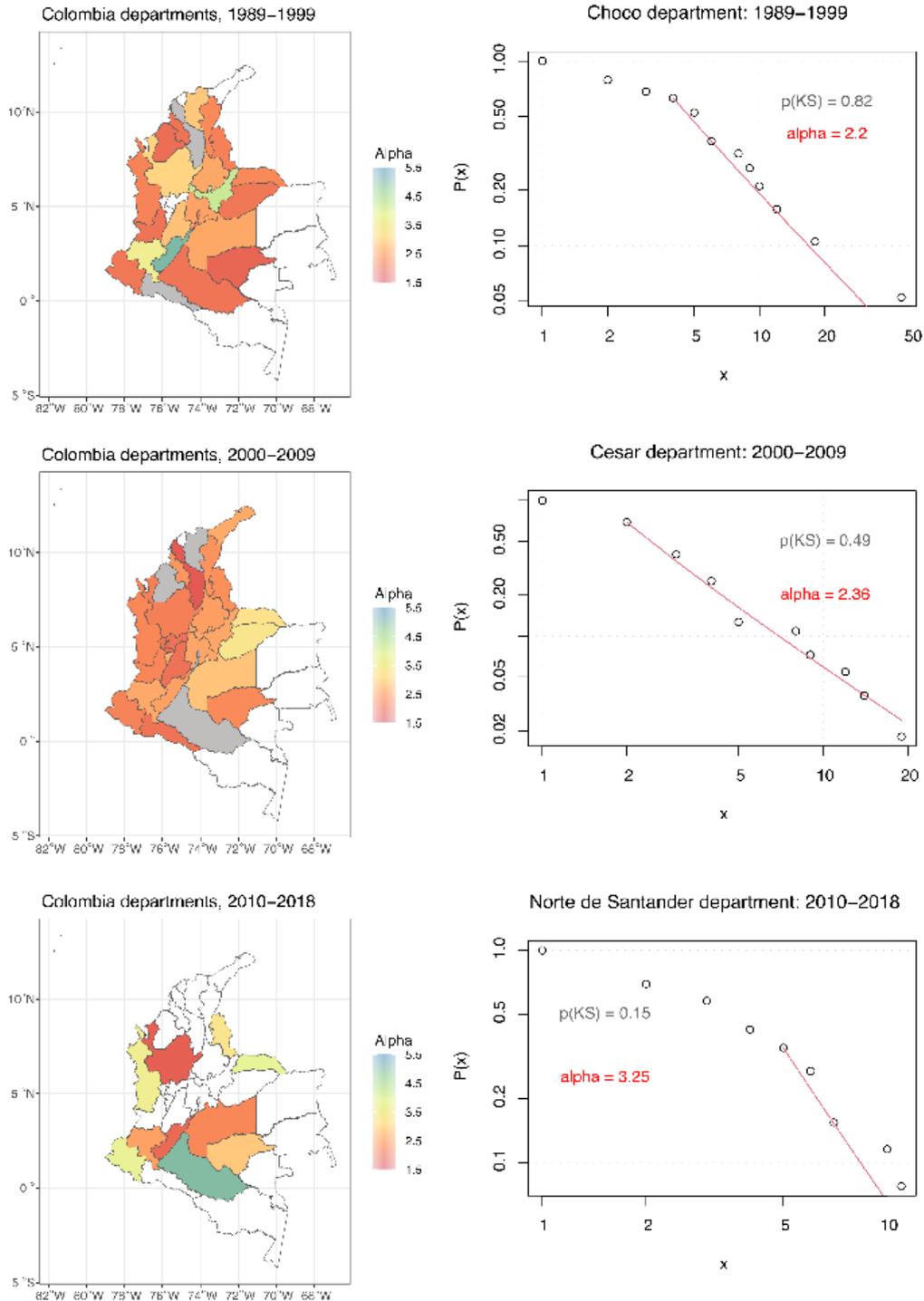

Figure 4: Results for the departments in Colombia obtained by fitting the data in the power-law distribution and calculating the Kolmogorov-Smirnov goodness of fit test. Left: Geographical distribution of exponent alpha, departments with P-value < 0.1 in grey, departments with no data in white. Right: CCDF for the selected departments.

# Discussion

We have shown that the data on conflict events for the armed conflict in Colombia and the individual segments of the conflict follow an approximate power-law distribution for conflict event severities with an exponent α varying broadly around the value 2.5. To the best of our knowledge, this is the first attempt to confirm the presence of the power-law distribution in armed conflict on such a fine-grained scale. While some of the spatial variation in the power-law exponent might be related to the random subsampling variation, we are confident that the further variation that we detect is not random and instead corresponds with the actor constellations within the individual conflict segments. This is validated by the results of our resampling analysis. Further details are provided in the Methods section.

The existence of such a robust pattern has several important consequences for the future analysis of armed conflict dynamics. First, our results show that detailed conflict analysis is a useful tool to understand how interconnected subconflicts form a larger conflict.

Second, the power-law distribution is a scale-free distribution meaning that conflict events are the results of the same generative process[8] and imply that there is a potential to have one theory explaining conflict events occurrences without distinguishing between large and small events.

Third, knowing that the underlying distribution of conflict events frequency and size fit power-law distribution allows for the inference of the missing fatalities in the data. This is particularly useful for conflict research as we know that conflict events, especially in remote areas, tend to be underreported. Bayesian modeling of different types of errors, for example, erroneous counting of fatalities or the use of rounding in official data, was successfully used to fill the gaps in the data on overall fatalities numbers in the American Indian War[44].

Fourth, the presence of the power-law distribution across different segments of the armed conflict in Colombia suggests the possibility of predicting the occurrence and the size of future conflict events simultaneously[6]. This is crucial from the practitioners' perspective as the impact of large events is distinctive from the small ones. Being able to anticipate a large conflict event in a specific region of Colombia can enable practitioners to better plan and, thus, more efficiently mitigate the impact of armed conflict on civilians.

Finally, we show that the fluctuations in the scaling exponent $\alpha$ correspond with changes in the actor constellations in line with the conflict actor fragmentation-coalescence model. The increase in the magnitude of the exponent, meaning fewer large conflict events, is associated with an increasing number of actors or a skewed strength ratio between the actors. Similarly, the decrease in the magnitude of the exponent depicts a higher frequency of large conflict events, which corresponds with a more balanced strength ratio. Future research should explore the possibility of using the $\alpha$ exponent as a proxy for the strength ratio of conflict actors more systematically and across the universe of armed conflicts.

# Methods

## Data

We obtain information on conflict events, their location, date, actors involved and the number of fatalities from the Uppsala Conflict Data Programme Georeferenced Event Dataset version 19.1 (UCDP GED)[37,38]. UCDP defines a conflict event as '[a]n incident where armed force was used by an organized actor against another organized [sic] actor, or against civilians, resulting in at least 1 direct death at a specific location and a specific date'[45].

For our analysis, we select only conflict events that have accurate enough location information available that events can be assigned to a specific department. This enables us to assign these events to specific departments. To achieve this, we rely on the UCDP variable called "where_prec" and include only events with values equal to or smaller than four. This eliminates 214 out of 3517 conflict events.

## Analysis

A power-law distribution describes the frequency of conflict events of a size $x$, as $p(x) = x^{-\alpha}/\zeta(\alpha, x_{\min})$, where $\alpha$ is a scaling exponent and $\zeta(\alpha, x_{\min})$ is a Hurwitz zeta function for discrete power-laws which is more appropriate for our type of data [8]. When plotted on a log-log plot where the x-axis is the logarithm of the event size and the y-axis is the logarithm of the cumulative probability $P(x)$ of that event being at least size $x$, the data produce a straight line with a slope $\alpha - 1$ [12] (N.B. all figures show P(x)). A power-law distribution is also often called scale-free. This means that increasing the scale or changing the unit in which $x$ is measured does not result in any change in the shape of the distribution[13,14].

We used the following procedure to fit our data to the power-law distribution. First, we estimated $\alpha$ and $x_{min}$ from our data to draw the ideal power-law distribution based on these parameters. Second, we calculated the Kolmogorov-Smirnov goodness of fit test and obtained the P-value via the bootstrap procedure with 5000 iterations. We accept that data follow approximate power-law distribution if the resulting P-value is equal to or greater than 0.1[10]. The complete results are reported in Tables 2, 3a, and 3b.

## Random sub-sampling

To ensure that the variation in the estimated value of the exponent α across the conflict segments has some meaning and does not arise entirely by chance, we draw random subsamples, estimate the exponent α for them and compare those estimates with the exponent values obtained for the individual conflict segments. More specifically, each conflict period $j$ has $N_j$ total events across Colombia, consisting of $n_{ij}$ events in each region $i$. We sample 50 times $n_{ij}$ points from $N_j$ to simulate the alternative composition of region $i$ in each period.

Afterwards, we fit each simulated region into the power-law distribution and below we report the values of the scaling exponent $\alpha$. We performed the Kolmogorov-Smirnov goodness of fit test to obtain the P-value via bootstrapping with 5000 iterations. We repeat that same process for all departments.

We include in our results presented below, only those values of $\alpha$ that have their corresponding P-values equal to or greater than 0.1. Results comprising all values of $\alpha$, including those with lower P-values, follow similar patterns. This applies to both regions and departments.

Figure 5 depicts how the values of $\alpha$ for the simulated alternative regions in the periods 1989-1999 and 2010-2018 remained relatively stable in comparison to the values of $\alpha$ estimated for the actual regions. The median values of the exponent for simulated regions during the period 2000-2009 tend to oscillate more than the other periods, but the variation in $\alpha$ is still lower compared to the estimated values for the actual regions in the same period. Table 4 further demonstrates that the simulated alternative regions have a lower level of variation in $\alpha$ compared to the actual data. The range of mean values for the simulated regions is smaller for each period than the range of actual values. Moreover, only some values of $\alpha$ fall within the range of 25$^{th}$ and 75$^{th}$ percentile or $\pm$ 1 standard deviation from the mean of the simulated regions for each studied period.

| Period | Region | Simulated results for alternative regions | | | | Results for regions | | Comparison | |
|---|---|---|---|---|---|---|---|---|---|
| | | Alpha mean | Alpha SD | Alpha 25th percentile | Alpha 75th percentile | Alpha | P-value (KS) | Within 25th-75th percentile | Within 1 SD |
| 1989 - 1999 | Region 1 | 2.49 | 0.37 | 2.23 | 2.72 | 2.43 | 0.06 | TRUE | TRUE |
| | Region 3 | 2.39 | 0.33 | 2.12 | 2.61 | 2.36 | 0.18 | TRUE | TRUE |
| | Region 4 | 2.5 | 0.59 | 2.12 | 2.71 | 3.03 | 0.67 | FALSE | TRUE |
| | Region 5 | 2.39 | 0.3 | 2.14 | 2.53 | 3.28 | 0.24 | FALSE | FALSE |
| | Region 6 | 2.42 | 0.36 | 2.12 | 2.62 | 2.51 | 0.59 | TRUE | TRUE |
| | Region 2 | 2.41 | 0.27 | 2.19 | 2.59 | 2.8 | 0.54 | FALSE | FALSE |
| | Region 7 | 2.47 | 0.41 | 2.24 | 2.68 | 1.92 | 0.46 | FALSE | FALSE |
| | Range of alpha | 2.39 - 2.50 | | | | 1.92 - 3.28 | | | |
| 2000 - 2009 | Region 1 | 2.49 | 0.2 | 2.38 | 2.57 | 2.17 | 0.77 | FALSE | FALSE |
| | Region 2 | 2.45 | 0.24 | 2.3 | 2.56 | 2.04 | 0.28 | FALSE | FALSE |
| | Region 3 | 2.29 | 0.23 | 2.14 | 2.4 | 2.27 | 0.22 | TRUE | TRUE |
| | Region 4 | 2.39 | 0.2 | 2.22 | 2.48 | 2.46 | 0.86 | TRUE | TRUE |
| | Region 5 | 2.47 | 0.22 | 2.32 | 2.67 | 2.29 | 0.49 | FALSE | TRUE |
| | Region 6 | 2.47 | 0.25 | 2.3 | 2.63 | 2.61 | 0.53 | TRUE | TRUE |
| | Region 7 | 2.48 | 0.22 | 2.32 | 2.61 | 2.39 | 0.25 | TRUE | TRUE |
| | Range of alpha | 2.29 - 2.49 | | | | 2.04 - 2.61 | | | |
| 2010 - 2018 | Region 1 | 2.28 | 0.58 | 1.87 | 2.52 | 2.45 | 0.81 | TRUE | TRUE |
| | Region 2 | 2.39 | 0.39 | 2.1 | 2.59 | 1.79 | 0.04 | FALSE | FALSE |
| | Region 4 | 2.5 | 0.41 | 2.22 | 2.71 | 2.35 | 0.27 | TRUE | TRUE |
| | Region 5 | 2.62 | 0.59 | 2.21 | 2.93 | 2.35 | 0.01 | TRUE | TRUE |
| | Region 6 | 2.57 | 0.58 | 2.27 | 2.8 | 5.34 | 0.79 | FALSE | FALSE |
| | Region 7 | 2.46 | 0.55 | 2.14 | 2.57 | 3.16 | 0.93 | FALSE | FALSE |
| | Region 3 | 2.47 | 0.64 | 2.06 | 2.68 | 3.25 | 0.13 | FALSE | FALSE |
| | Range of alpha | 2.28 - 2.62 | | | | 1.79 - 5.34 | | | |

Table 4: Comparison of the results obtained from the bootstrapping procedure by simulating alternative regions for each of the studied periods. The last two columns on the right side of the table compare the values of alpha to the simulated results.

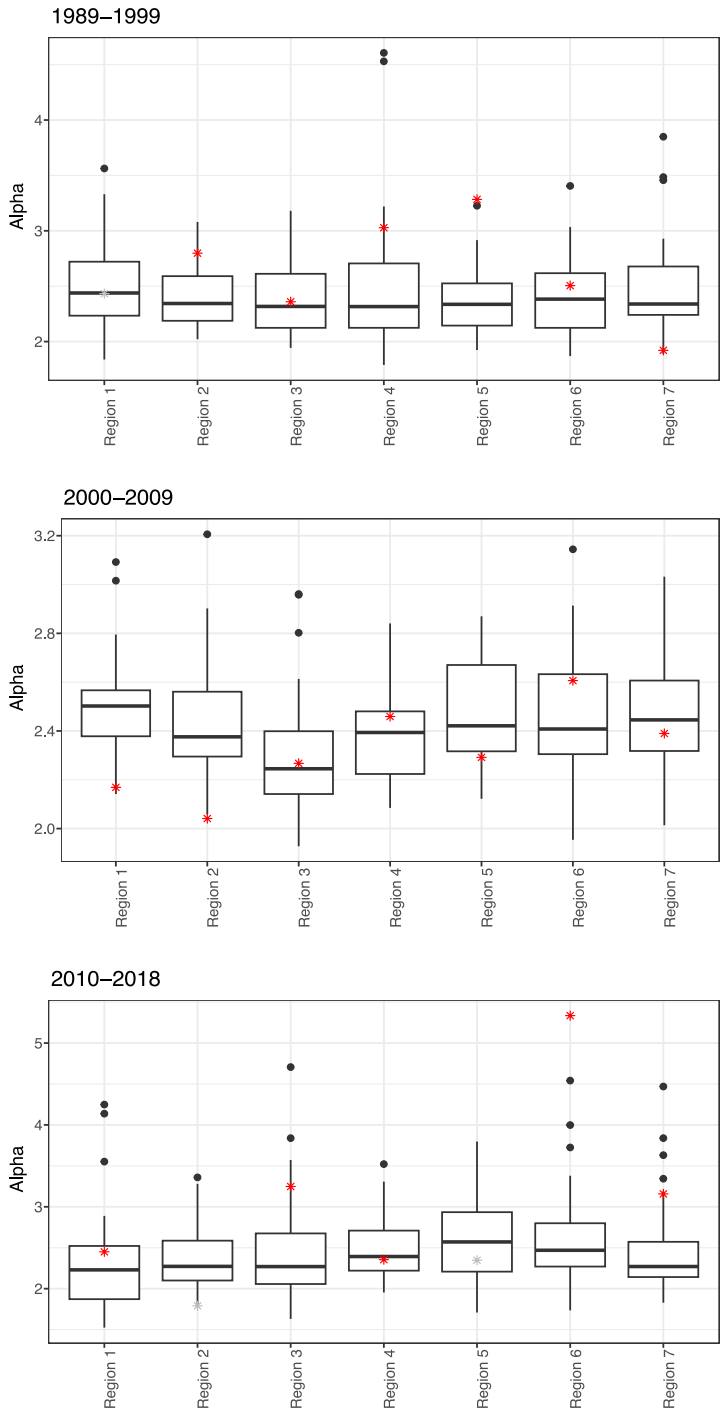

Figure 5: Results obtained from the bootstrapping procedure by simulating alternative regions for each of the studied periods. The boxplot displays the median, 25th and 75th quantiles of the alpha values for the simulated regions. The alpha values estimated for the actual regions are depicted by stars, with corresponding P-value ≥ 0.1 in red, otherwise in grey.

Figure 6 and table 5 show that the conflict segments based on departments exhibit similar patterns. The mean and median values for the simulated alternative departments tend to vary within a narrower range compared to the estimated values for the actual departments. As in the case of regions, some of the values of $\alpha$ estimated for the actual departments fall outside the 15th and 75th percentile and are greater or smaller than the mean of the simulated departments $\pm$ 1 standard deviation. This supports our claim that the $\alpha$ value for an actual department does generally contain information about its specific conflict dynamics. Such information is not present in a randomized version. Hence the $\alpha$ values that we report for the conflict on the fine-grained scale, are meaningful.

| Period | Department | Simulated results for alternative departments | | | | Results for departments | | Comparison | |
|---|---|---|---|---|---|---|---|---|---|
| | | Alpha mean | Alpha SD | Alpha 25th percentile | Alpha 75th percentile | Alpha | P-value (KS) | Within 25th-75th percentile | Within 1 SD |
| 1989-1999 | Antioquia department | 2.47 | 0.6 | 2.14 | 2.59 | 2.98 | 0.77 | FALSE | TRUE |
| | Arauca department | 2.52 | 0.54 | 2.16 | 2.73 | 2.26 | 0.21 | TRUE | TRUE |
| | Bogota department | 2.65 | 0.68 | 2.07 | 2.95 | 2.2 | 0.12 | TRUE | TRUE |
| | Boyaca department | 2.69 | 0.55 | 2.2 | 3.01 | 4.19 | 0.14 | FALSE | FALSE |
| | Cordoba department | 2.53 | 0.52 | 2.06 | 2.86 | 1.97 | 0.33 | FALSE | FALSE |
| | Caqueta department | 2.58 | 0.63 | 2.09 | 3.04 | 2.03 | 0.18 | FALSE | TRUE |
| | Cauca department | 2.53 | 0.51 | 2.23 | 2.71 | 3.58 | 0.7 | FALSE | FALSE |
| | Cesar department | 2.76 | 0.75 | 2.21 | 3.16 | 2.26 | 0.37 | TRUE | TRUE |
| | Choco department | 2.5 | 0.6 | 2 | 2.88 | 2.2 | 0.82 | TRUE | TRUE |
| | Cundinamarca d. | 2.38 | 0.53 | 2.03 | 2.71 | 2.51 | 0.27 | TRUE | TRUE |
| | Guaviare department | 2.41 | 0.41 | 2.12 | 2.68 | 1.88 | 0.61 | FALSE | FALSE |
| | Huila department | 2.7 | 0.86 | 2.12 | 2.82 | 4.99 | 0.47 | FALSE | FALSE |
| | Magdalena d. | 2.41 | 0.49 | 2.04 | 2.64 | 2.86 | 0.91 | FALSE | TRUE |
| | Meta department | 2.46 | 0.46 | 2.13 | 2.82 | 2.55 | 0.72 | TRUE | TRUE |
| | Narino department | 2.65 | 0.63 | 2.25 | 2.79 | 2.07 | 0.19 | FALSE | TRUE |
| | Norte de Santander d. | 2.51 | 0.44 | 2.16 | 2.78 | 2.17 | 0.14 | TRUE | TRUE |
| | Sucre department | 2.61 | 0.71 | 1.98 | 2.88 | 2.02 | 0.51 | TRUE | TRUE |
| | Tolima department | 2.56 | 0.64 | 2.15 | 2.95 | 2.8 | 0.77 | TRUE | TRUE |
| | Valle del Cauca d. | 2.56 | 0.62 | 2.11 | 2.84 | 2.02 | 0.95 | FALSE | TRUE |
| | Bolivar department | 2.62 | 0.72 | 2.15 | 2.89 | 1.76 | 0.01 | FALSE | FALSE |
| | Casanare department | 2.46 | 0.76 | 1.93 | 2.62 | 2.07 | 0.59 | TRUE | TRUE |
| | Putumayo department | 2.43 | 0.64 | 1.89 | 2.75 | 1.81 | 0.09 | FALSE | TRUE |
| | Santander department | 2.57 | 0.57 | 2.19 | 2.75 | 2.55 | 0.75 | TRUE | TRUE |
| | Range of alpha | 2.38 - 2.76 | | | | 1.76 - 4.99 | | | |

Table 5a (part 1): Comparison of the results obtained from the bootstrapping procedure by simulating alternative departments for each of the studied periods. The last two columns on the right side of the table compare the values of alpha to the simulated results.

| Period | Department | Simulated results for alternative departments | | | | Results for departments | | Comparison | |
|---|---|---|---|---|---|---|---|---|---|
| | | Alpha mean | Alpha SD | Alpha 25th percentile | Alpha 75th percentile | Alpha | P-value (KS) | Within 25th-75th percentile | Within 1 SD |
| 2000 - 2009 | Antioquia department | 2.46 | 0.23 | 2.33 | 2.61 | 2.16 | 0.36 | FALSE | FALSE |
| | Arauca department | 2.43 | 0.47 | 2.08 | 2.62 | 3.15 | 0.91 | FALSE | FALSE |
| | Bogota department | 2.56 | 0.71 | 2.04 | 2.86 | 1.72 | 0.05 | FALSE | FALSE |
| | Bolivar department | 2.46 | 0.36 | 2.17 | 2.69 | 1.76 | 0.18 | FALSE | FALSE |
| | Boyaca department | 2.56 | 0.66 | 2.12 | 2.77 | 2.55 | 0.12 | TRUE | TRUE |
| | Cordoba department | 2.32 | 0.49 | 1.97 | 2.65 | 1.64 | 0.07 | FALSE | FALSE |
| | Caldas department | 2.37 | 0.45 | 2.05 | 2.52 | 2.01 | 0.22 | FALSE | TRUE |
| | Caqueta department | 2.36 | 0.28 | 2.18 | 2.43 | 2.86 | 0.05 | FALSE | FALSE |
| | Casanare department | 2.63 | 0.73 | 2.15 | 2.69 | 3.29 | 0.94 | FALSE | TRUE |
| | Cauca department | 2.46 | 0.43 | 2.18 | 2.61 | 2.56 | 0.64 | TRUE | TRUE |
| | Cesar department | 2.48 | 0.55 | 2.1 | 2.67 | 2.36 | 0.49 | TRUE | TRUE |
| | Choco department | 2.43 | 0.49 | 2.12 | 2.66 | 2.23 | 0.49 | TRUE | TRUE |
| | Cundinamarca d. | 2.42 | 0.44 | 2.09 | 2.65 | 2.53 | 0.65 | TRUE | TRUE |
| | Guaviare department | 2.46 | 0.5 | 2.16 | 2.67 | 2.27 | 0.56 | TRUE | TRUE |
| | Huila department | 2.46 | 0.51 | 2.16 | 2.55 | 2.49 | 0.85 | TRUE | TRUE |
| | Magdalena d. | 2.45 | 0.42 | 2.18 | 2.63 | 2.25 | 0.04 | TRUE | TRUE |
| | Meta department | 2.42 | 0.27 | 2.26 | 2.59 | 2.82 | 0.6 | FALSE | FALSE |
| | Narino department | 2.42 | 0.41 | 2.14 | 2.55 | 2.19 | 0.63 | TRUE | TRUE |
| | Norte de Santander d. | 2.43 | 0.43 | 2.17 | 2.57 | 2.32 | 0.15 | TRUE | TRUE |
| | Putumayo department | 2.45 | 0.59 | 2.12 | 2.67 | 2.03 | 0.68 | FALSE | TRUE |
| | Quindio department | 2.42 | 0.58 | 1.99 | 2.81 | 2.65 | 0.34 | TRUE | TRUE |
| | Risaralda department | 2.5 | 0.49 | 2.07 | 2.87 | 2.05 | 0.37 | FALSE | TRUE |
| | Santander department | 2.44 | 0.44 | 2.14 | 2.63 | 2.46 | 0.72 | TRUE | TRUE |
| | Sucre department | 2.57 | 0.69 | 2.13 | 2.67 | 2.39 | 0.1 | TRUE | TRUE |
| | Tolima department | 2.4 | 0.33 | 2.17 | 2.61 | 2 | 0.27 | FALSE | FALSE |
| | Valle del Cauca d. | 2.34 | 0.25 | 2.17 | 2.45 | 2.23 | 0.73 | TRUE | TRUE |
| | La Guajira d. | 2.77 | 0.72 | 2.22 | 3.18 | 2.58 | 0.94 | TRUE | TRUE |
| | **Range of alpha** | 2.32 - 2.77 | | | | 1.64 - 3.29 | | | |

Table 5b (part 2): Comparison of the results obtained from the bootstrapping procedure by simulating alternative departments for each of the studied periods. The last two columns on the right side of the table compare the values of alpha to the simulated results.

| Period | Department | Simulated results for alternative departments | | | | Results for departments | | Comparison | |
|---|---|---|---|---|---|---|---|---|---|
| | | Alpha mean | Alpha SD | Alpha 25th percentile | Alpha 75th percentile | Alpha | P-value (KS) | Within 25th-75th percentile | Within 1 SD |
| 2010 - 2018 | Antioquia department | 2.45 | 0.53 | 2.12 | 2.66 | 2.16 | 0.36 | TRUE | TRUE |
| | Arauca department | 2.59 | 0.7 | 2.06 | 2.92 | 3.15 | 0.91 | FALSE | TRUE |
| | Cauca department | 2.5 | 0.53 | 2.12 | 2.6 | 2.56 | 0.64 | TRUE | TRUE |
| | Choco department | 2.51 | 0.63 | 2.02 | 2.7 | 2.23 | 0.49 | TRUE | TRUE |
| | Guaviare department | 2.36 | 0.48 | 1.94 | 2.69 | 2.27 | 0.56 | TRUE | TRUE |
| | Huila department | 2.54 | 0.56 | 2.03 | 2.93 | 2.49 | 0.85 | TRUE | TRUE |
| | Meta department | 2.56 | 0.59 | 2.09 | 2.79 | 2.82 | 0.6 | FALSE | TRUE |
| | Narino department | 2.62 | 0.71 | 2.12 | 2.96 | 2.19 | 0.63 | TRUE | TRUE |
| | Norte de Santander d. | 2.4 | 0.49 | 2.08 | 2.56 | 2.32 | 0.15 | TRUE | TRUE |
| | Caqueta department | 2.66 | 0.66 | 2.19 | 3.26 | 2.86 | 0.05 | TRUE | TRUE |
| | **Range of alpha** | 2.36 - 2.66 | | | | 2.16 - 3.15 | | | |

Table 5c (part 3): Comparison of the results obtained from the bootstrapping procedure by simulating alternative departments for each of the studied periods. The last two columns on the right side of the table compare the values of alpha to the simulated results.

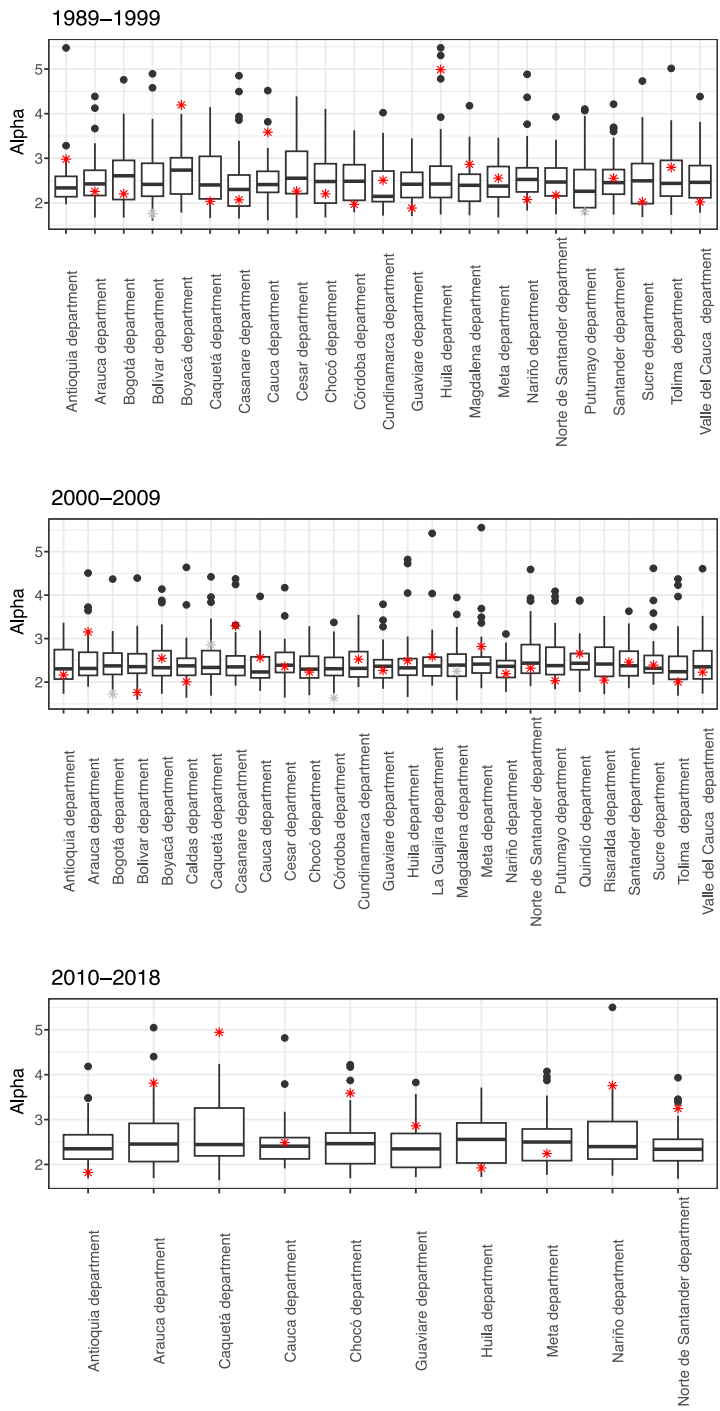

Figure 6: Results obtained from the bootstrapping procedure by simulating alternative departments for each of the studied periods. The boxplot displays the median, 25th and 75th quantiles of the alpha values for the simulated departments. The alpha values estimated for the actual departments are depicted by stars, with corresponding P-value ≥ 0.1 in red, otherwise in grey.

## Code and data availability

Data and R scripts are available at: https://zenodo.org/records/10159421.

We carried out all calculations using R version 4.3.1 ("Beagle Scouts") and R package poweRlaw 0.70.6 [46].

## References

1. Richardson, L. F. Variation of the frequency of fatal quarrels with magnitude. *Journal of the American Statistical Association* **43**, 523–546 (1948).
2. Richardson, L. F. *Statistics of Deadly Quarrels*. vol. 10 (Boxwood Press, 1960).
3. Biggs, M. Strikes as forest fires: Chicago and Paris in the late nineteenth century. *American Journal of Sociology* **110**, 1684–1714 (2005).
4. Cederman, L. E. Modeling the size of wars: From billiard balls to sandpiles. *American Political Science Review* **97**, 135–150 (2003).
5. Bohorquez, J. C., Gourley, S., Dixon, A. R., Spagat, M. & Johnson, N. F. Common ecology quantifies human insurgency. *Nature* **462**, 911–914 (2009).
6. Spagat, M., Johnson, N. F. & Weez, S. V. Fundamental patterns and predictions of event size distributions in modern wars and terrorist campaigns. *PLoS ONE* **13**, 1–13 (2018).
7. Clauset, A. & Young, M. Scale Invariance in Global Terrorism. **87131**, (2005).
8. Clauset, a., Young, M. & Gleditsch, K. S. On the Frequency of Severe Terrorist Events. *Journal of Conflict Resolution* **51**, 58–87 (2007).
9. Idler, A. *Borderland Battles: Violence, Crime, and Governance at the Edges of Colombia's War*. (Oxford University Press, 2019).
10. Clauset, A., Shalizi, C. R. & Newman, M. E. J. Power-law distributions in empirical data. *SIAM Review* **51**, 661–703 (2009).
11. Clauset, A. On the frequency and severity of interstate wars. in *Lewis Fry Richardson: His Intellectual Legacy and Influence in the Social Sciences* 113–127 (Springer Nature, 2020).
12. Friedman, J. A. Using Power Laws to Estimate Conflict Size. *Journal of Conflict Resolution* **59**, 1216–1241 (2015).
13. Newman, M. E. J. Power laws, Pareto distributions and Zipf's law. *Contemporary Physics* **46**, 323–351 (2005).
14. Kumamoto, S.-I. & Kamihigashi, T. Power Laws in Stochastic Processes for Social Phenomena: An Introductory Review. *Frontiers in Physics* **6**, 1–17 (2018).
15. Malamud, B. D., Morein, G. & Turcotte, D. L. Forest fires: An example of self-organized critical behavior. *Science* **281**, 1840–1842 (1998).
16. Brunk, G. G. Why Do Societies Collapse? *Journal of Theoretical Politics* **14**, 195–230 (2002).
17. Bak, P. & Chan, K. Self-Organized Criticality. *Scientific American* **264**, 46–53 (1991).
18. Ruszczycki, B., Burnett, B., Zhao, Z. & Johnson, N. Relating the microscopic rules in coalescence-fragmentation models to the cluster-size distribution. *European Physical Journal B* **72**, 289–302 (2009).

19. Coulson, S. G. Lanchester modelling of intelligence in combat. *IMA Journal of Management Mathematics* **30**, 149–164 (2017).
20. Kress, M., Caulkins, J. P., Feichtinger, G., Grass, D. & Seidl, A. Lanchester model for three-way combat. *European Journal of Operational Research* **264**, 46–54 (2018).
21. Idler, A. The Logic of Illicit Flows in Armed Conflict: Explaining Variation in Violent Nonstate Group Interactions in Colombia. *World Politics* 1–42 (2020).
22. Avilés, W. Institutions, Military Policy, and Human Rights in Colombia. *Latin American Perspectives* **28**, 31–55 (2001).
23. Peceny, M. & Durnan, M. The FARC's Best Friend: U.S. Antidrug Policies and the Deepening of Colombia's Civil War in the 1990s. *Lat. Am. polit. soc.* **48**, 95–116 (2006).
24. Thomson, A. The credible commitment problem and multiple armed groups: FARC perceptions of insecurity during disarmament in the Colombian peace process. *Conflict, Security & Development* **20**, 497–517 (2020).
25. Safford, F. & Palacios, M. *Colombia: Fragmented Land, Divided Society*. (Oxford University Press, 2002).
26. Idler, A. Local Peace Processes in Colombia. in *Local Peace Processes* (ed. Political Settlements Research Programme) (British Academy, 2021).
27. Sacquet, T. J. Colombian Military Forces. *Veritas* **2**, (2006).
28. ESISC Team. *Colombia: an overview of the Farc's military structure*. http://www.esisc.org/publications/briefings/colombia-an-overview-of-the-farcs-military-structure (2010).
29. Tras 50 años de guerra, las Farc están débiles más no derrotadas. *El País* (2014).
30. GameChangers 2017: Is Colombia's FARC Really Gone? *InSight Crime* https://insightcrime.org/news/analysis/gamechangers-2017-is-colombia-farc-really-gone/ (2018).
31. Ejército Nacional de Colombia. #Atención No olvides seguir las cuentas de Twitter de nuestras divisiones en toda #Colombia #VestidosdeHonor. *Twitter* https://twitter.com/col_ejercito/status/650042891046363138 (2015).
32. Nieto Rojas, J. H. Policía Nacional de Colombia: Informe de gestión institucional 2016. (2016).
33. Cabrera Nossa, I. A., Echandía Castilla, C., Cabrera Nossa, I. A. & Echandía Castilla, C. Retos institucionales y no institucionales para el partido Fuerza Alternativa Revolucionaria del Común (FARC) en las elecciones legislativas de 2018. *Estudios Políticos* 92–121 (2019) doi:10.17533/udea.espo.n56a05.
34. Colombia Constitution. (1991).
35. Gutiérrez Sanín, F. Instituciones y territorio. La descentralización en Colombia. in *25 años de la descentralización en Colombia* (ed. Konrad Adenauer Stiftung) 11–54 (Konrad Adenauer Stiftung, 2010).
36. Idler, A. & Paladini Adell, B. When Peace Implies Engaging the "Terrorist": Peacebuilding in Colombia through Transforming Political Violence and Terrorism. in *The Nexus between Terrorism Studies and Peace and Conflict Studies* (eds. Tellidis, Y. & Toros, H.) (Routledge, 2015).

37. Croicu, M. & Sundberg, R. UCDP Georeferenced Event Dataset Codebook Quick Start Guide : 1–38 (2018).
38. Pettersson, T. & Eck, K. Organized violence, 1989–2017. *Journal of Peace Research* **55**, 535–547 (2018).
39. Beardsley, K., Gleditsch, K. S. & Lo, N. Roving Bandits? The Geographical Evolution of African Armed Conflicts. *International Studies Quarterly* **59**, 503–516 (2015).
40. Wood, R. M. & Kathman, J. D. Competing for the Crown: Inter-rebel Competition and Civilian Targeting in Civil War. *Political Research Quarterly* **68**, 167–179 (2015).
41. Ito, G. & Hinkkainen Elliott, K. Battle Diffusion Matters: Examining the Impact of Microdynamics of Fighting on Conflict Termination. *Journal of Conflict Resolution* **64**, 871–902 (2020).
42. Idler, A. & Tkacova, K. Conflict shapes in flux: Explaining spatial shift in conflict-related violence. *International Political Science Review* 01925121231177445 (2023).
43. Eccarius-Kelly, V. Surreptitious Lifelines: A Structural Analysis of the FARC and the PKK. *Terrorism and Political Violence* **24**, 235–258 (2012).
44. Gillespie, C. S. Estimating the number of casualties in the American Indian war: A Bayesian analysis using the power law distribution. *The Annals of Applied Statistics* **11**, 2357–2374 (2017).
45. Högbladh, S. *UCDP GED Codebook version 20.1.* https://ucdp.uu.se/downloads/ged/ged201.pdf (2020).
46. Gillespie, C. S. Fitting Heavy Tailed Distributions: The poweRlaw Package. *Journal of Statistical Software* **64**, 1–16 (2015).

**Acknowledgments**

This study is based upon work supported by the Minerva Research Initiative in partnership with the Air Force Office of Scientific Research under award number FA9550-22-1-0338 awarded to AI as Principal Investigator and NJ and EL as Co-Investigators; by the Arts & Humanities Research Council and the Economic & Social Research Council through the Partnership for Conflict, Crime & Security Research under Grant Ref: AH/P005446/1 awarded to AI as Principal Investigator; and by the Arts & Humanities Research Council under Grant Ref: AH/V011375/1 awarded to AI as Principal Investigator. NJ is also supported by U.S. Air Force Office of Scientific Research awards FA9550-20-1-0382 and FA9550-20-1-0383.

**Author contributions statement**

KT, AI, NJ, and EL conceived and designed the experiments and wrote the paper. KT and AI contributed materials/analysis tools. KT performed the experiments and analyzed the data.

**Supplementary material**

All data used, generated, or analyzed during this study are included in this article or supplementary files. Data and R scripts are also available at: https://zenodo.org/records/10159421.

**Competing interests**

The authors declare no competing interests.